\documentclass{Interspeech}

\interspeechcameraready 

\usepackage{multirow}

\title{First Steps Towards Voice Anonymization for Code-Switching Speech}

\author{Sarina}{Meyer}
\author{Ekaterina}{Kolos}
\author{Ngoc Thang}{Vu}

\affiliation[nocounter]{Institute for Natural Language Processing}{University of Stuttgart}{Germany}
\email{sarina.meyer@ims.uni-stuttgart.de}

\newcommand{\mcadams}{B2}
\newcommand{\xiaoxiao}{SSL}
\newcommand{\our}{$\text{GAN}_{\text{multi}}$}
\newcommand{\seame}{\textsc{Seame}}
\newcommand{\miami}{\textsc{Miami}}
\usepackage{comment}

\begin{document}

\maketitle

\begin{abstract}
    The goal of voice anonymization is to modify an audio such that the true identity of its speaker is hidden. Research on this task is typically limited to the same English read speech datasets, thus the efficacy of current methods for other types of speech data remains unknown. In this paper, we present the first investigation of voice anonymization for the multilingual phenomenon of code-switching speech. We prepare two corpora for this task and propose adaptations to a multilingual anonymization model to make it applicable for code-switching speech. By testing the anonymization performance of this and two language-independent methods on the datasets, we find that only the multilingual system performs well in terms of privacy and utility preservation. Furthermore, we observe challenges in performing utility evaluations on this data because of its spontaneous character and the limited code-switching support by the multilingual speech recognition model.
\end{abstract}

\section{Introduction}
Voice anonymization (VA) refers to the task of hiding the identity of a speaker in an audio. Common approaches modify the audio using signal processing \cite{mcadams}, voice conversion \cite{miao_2022_language, nac, shamsabadi2022differentially}, or cascade systems of speech recognition (ASR) and text-to-speech (TTS) \cite{b3}. The main goal, as defined by the  Voice Privacy Challenges (VPC) \cite{introducing_vpc, vpc22_taslp, vpc24_eval_plan}, is to protect the speaker's privacy from being identified by a speaker verification (ASV) system while keeping the utility for a downstream ASR task.

Most VA systems are evaluated using the procedures proposed by the VPC. While the exact test conditions and metrics differ in each edition, the evaluation datasets remain constant at subsets of LibriSpeech \cite{librispeech} and 
VCTK  \cite{vctk}.\footnote{VCTK was replaced by IEMOCAP \cite{iemocap} in VPC 2024 \cite{vpc24_eval_plan} to measure emotional preservation but not privacy or linguistic utility.}
These datasets consist of single-speaker read speech, are monolingual in English, and recorded in noise-free environments. Thus, VA approaches are mainly evaluated on data that differs greatly from speech found in the wild such as spontaneous conversations. One aspect of many conversations across the world is that they are not, or not only, in English. While there have been previous approaches to explore VA for different languages, by developing multilingual \cite{meyer24_multilingual, yao2024musa} and language-independent \cite{miao_2022_language, miao_2022_analyzing} methods, to this date, the presence of several languages within one recording has not yet been investigated. This commonly occurs in multilingual communities as \textit{code-switching}, in which a speaker alternates between two or more languages within a conversation, sentence or word. 
Code-switching is a phenomenon that has been found challenging for speech processing topics in the past \cite{mustafa2022code}, yet its effects on VA are so far unknown.

Thus, in this paper, we present to our knowledge the first investigation of VA in the context of code-switching. 
Specifically, we start by analyzing how well current VA methods deal with code-switching speech. While the way speakers code-switch differs between individuals \cite{auer1999cs} and might be preferred to be changed during anonymization, as a starting point, we assume that the code-switching behavior of the original speaker should be preserved. 
We prepare two existing code-switching datasets from Mandarin-English and Spanish-English communities for their applicability in anonymization and evaluate the performances of two language-independent VA models and one multilingual one that we adapt for code-switching. 
We find the anonymization of the multilingual model to be generally successful, achieving high privacy scores and only little degradation in utility, whereas the language-independent models offer almost no privacy protection already on monolingual audios.
At the same time, the experiments reveal special challenges for the utility evaluation of this kind of data. 
While Whisper, being a powerful multilingual ASR model, is capable of keeping the speakers' code-switching behavior to some extent, its performance on the monolingual utterances of these datasets with 20-30\% mixed error rate (MER) is substantially worse than the 1.5-3\% achieved on the standard VPC datasets. Our analysis reveals that this is mainly due to the code-switching typical spontaneous nature and audio quality of the data. We observe a further drop to $\sim39\%$ MER for code-switched utterances and find that the original code-switching behavior is significantly altered and reduced during speech recognition towards more monolingual transcriptions, due to omissions and translations.  
We conclude that the evaluation of VA needs to be adapted to the nature of the data that is being anonymized. Thus, there is a need for a more diverse evaluation framework that supports a variety of speech phenomena, including code-switching. 
We publish our code and data online to encourage more research in this field\footnote{\url{https://github.com/DigitalPhonetics/speaker-anonymization}}.

\section{Methods}

\subsection{Data}
\begin{table}[t]
    \centering
    \caption{Number of speakers and utterances of the \seame{} and \miami{} datasets, divided into female (F) and male (M) subsets, language settings (EN, ZH/ES, CS) and ASV sets (enroll, trial).}
    \resizebox{\columnwidth}{!}{
    \begin{tabular}{cc|r|rr|rr|rr}
        \toprule
        \multicolumn{2}{c|}{\textbf{\seame{}}} & & \multicolumn{2}{c|}{\textbf{\#utts EN}} & \multicolumn{2}{c|}{\textbf{\#utts ZH}} & \multicolumn{2}{c}{\textbf{\#utts CS}} \\
                               &   & \#spk & Enroll & Trial & Enroll & Trial & Enroll & Trial \\
        \midrule
        \multirow{2}{*}{Train} & F & 28    & \multicolumn{2}{c|}{1,632} & \multicolumn{2}{c|}{3,792} & \multicolumn{2}{c}{8,646} \\
                               & M & 26    & \multicolumn{2}{c|}{1,326} & \multicolumn{2}{c|}{2,824} & \multicolumn{2}{c}{9,942} \\
        \midrule
        \multirow{2}{*}{Dev}   & F & 10    & 62     & 311   & 66     & 994   & 91     & 3,261 \\
                               & M & 10    & 68     & 1,165  & 70     & 685   & 90     & 3,211 \\
        \midrule
        \multirow{2}{*}{Test}  & F & 12    & 86     & 472   & 86     & 983   & 114    & 5,246 \\
                               & M & 8     & 54     & 1,121  & 53     & 572   & 75     & 2,317 \\
        \bottomrule
        \toprule
        \multicolumn{2}{c|}{\textbf{\miami{}}} & & \multicolumn{2}{c|}{\textbf{\#utts EN}} & \multicolumn{2}{c|}{\textbf{\#utts ES}} & \multicolumn{2}{c}{\textbf{\#utts CS}} \\
                               &   & \#spk & Enroll & Trial & Enroll & Trial & Enroll & Trial \\
        \midrule
        \multirow{2}{*}{Test}  & F & 17    & 100    & 2,442  & 101    & 884   & 139    & 425 \\
                               & M & 8     & 55     & 942   & 56     & 362   &  68    & 201\\
        \bottomrule
    \end{tabular}
    }
    \label{tab:datasets}
\end{table}

We perform all experiments on two datasets, \seame{} (Mandarin-English) and \miami{} (Spanish-English). 
They contain two types of code-switching: Inter-sentential, in which the language switch happens between utterances, and intra-sentential with language switches within an utterance. 
We split the data into single utterances and denote only intra-sentential cases as \textit{code-switching} and everything else as \textit{monolingual}. Thus, we have three \textit{language settings} per dataset:  \textit{English} (EN), \textit{Mandarin} (ZH) and \textit{code-switching} (CS) in \seame{}, and \textit{English}, \textit{Spanish} (ES) and \textit{code-switching} in \miami{}. 

\textbf{\seame{}} \cite{seame} consists of 297 conversational and interview recordings of 156 Singaporean and Malaysian speakers using Mandarin-English code-switching. We select only the interview part of the data, 
comprising 94 speakers and 210 recordings. 20 speakers each are selected for the development and testing of the anonymization, while we reserve the remaining 54 speakers for training or finetuning of an ASV attacker.

Bangor Miami\footnote{\url{https://biling.talkbank.org/access/Bangor/Miami.html}} (shortened as \textbf{\miami{}}) consists of Spanish-English code-switching utterances of 56 conversational recordings of 84 speakers in Miami (USA). We exclude 15 recordings that contain only the speech of one specific speaker (María) and further exclude all speakers that have less than 20 utterances in each language setting. The remaining dataset comprises 25 speakers which we all use for testing. 

Each dataset is split into three parts according to the annotated language setting, with their statistics shown in Table \ref{tab:datasets}. 
We exclude utterances that overlap between speakers and remove annotations and discourse particles from transcriptions.
Short utterances of less than two seconds in duration are concatenated with other short utterances of the same speaker and similar loudness levels. In the end, all utterances in \seame{} have a duration between 2 and 42 seconds, with 130 to 1550 utterances per speaker. In \miami{}, utterances range between 2 to 21 seconds, with 110 to 400 utterances per speaker.   
We further split the development and test data into enrollment and trial subsets for ASV computation, with 4 to 10 utterances per speaker and language setting for enrollment and the remaining for trial.

These datasets are in several ways different to the datasets usually used for evaluating VA, namely LibriSpeech and VCTK. First, they are spontaneous and conversational speech. Utterances can be incomplete and ungrammatical, and contain repetitions, hesitations, or discourse particles. Moreover, being in a multilingual context, speakers might use words of other languages than the two primary ones and can have stronger accents. Finally, \miami{} has not been recorded in labs, thus recordings differ in audio quality. This includes noise and background speakers, as well as microphone differences.

\subsection{Voice anonymization}
We select three VA systems based on their ability to process several languages and their availability of open-source code. 

\textbf{\mcadams{}} \cite{mcadams, vpc24_eval_plan} is a parameter-free technique based on signal processing and is known as baseline B2 of the VPC. We use the code of the VPC 2024\footnote{\url{https://github.com/Voice-Privacy-Challenge/Voice-Privacy-Challenge-2024}}. It extracts pole positions from the input signal using linear predictive coding and then shifts these positions using the McAdams coefficient \cite{mcadams_coefficients}. 
This method is generally seen as a weak baseline with low privacy protection but has the advantage of being naturally language-independent.

\textbf{\xiaoxiao{}}\footnote{\url{https://github.com/nii-yamagishilab/SSL-SAS}} \cite{miao_2022_language, miao_2022_analyzing} is a neural voice anonymization method specifically designed to be language-independent. From an input signal, it extracts the linguistic content as soft content units using HuBERT \cite{hubert}, F0 information using YAAPT \cite{yaapt}, and the speaker embedding using an ECAPA-TDNN model \cite{ecapa}. The latter is anonymized by replacing it with an average of several speaker embeddings sampled from an external pool of speakers. The extracted and modified information is converted back into speech using a HiFiGAN vocoder \cite{hifigan}. 

\textbf{\our{}}\footnote{\url{https://github.com/DigitalPhonetics/speaker-anonymization}} is a multilingual version of baseline B3 of the VPC 2024, as proposed in our previous work \cite{meyer24_multilingual}. It is an ASR-TTS cascade system with a GAN-based \cite{gan} anonymization technique. Whisper \cite{whisper} is used to recognize the linguistic content as a transcript, and to identify the language of the input if a language has not been specified. The speaker information is anonymized by sampling an artificial ECAPA-TDNN-like embedding from a GAN. An audio is synthesized using a multilingual FastSpeech2-based TTS \cite{ren2021fastspeech} and a HiFiGAN vocoder, as provided in IMS Toucan \cite{toucan}. The system has the option to preserve prosodic information during anonymization, however, we disabled this functionality because it would produce unreliable pitch estimates for these datasets. Instead, we rely on the prosody estimation of the TTS, as it has been done in a previous version of this model \cite{meyer2023_gan}. Since the system expects only one language per audio, we extend this multilingual version to a code-switching variant. For this, we update the TTS and HiFiGAN models to their latest versions\footnote{\url{https://github.com/DigitalPhonetics/IMS-Toucan/releases/tag/v3.0}} because of their improved multilingual support \cite{lux2024massive}. During synthesis, a phonemizer and language embedding is chosen depending on the input language. For code-switching, we first detect the language of each word in the recognized transcript using either the dragonmapper tool\footnote{\url{https://github.com/tsroten/dragonmapper}} to recognize Chinese characters (\seame{}) or a BERT-based language identifier\footnote{\url{https://huggingface.co/sagorsarker/codeswitch-spaeng-lid-lince}} for Spanish-English code-switching (\miami{}). We then phonemize each word according to its language and concatenate the phonemized representations of all words. Lastly, we adapt the code such that the language embedding is chosen per phone instead of once per utterance.

\subsection{Evaluation} \label{subsec:evaluation}
The anonymization is evaluated using the framework of the VPC 2024, which is based on \cite{voicepat}. Each utterance is anonymized towards a different target speaker. 
For privacy, an ASV model is trained as \textit{semi-informed}, i.e., on data that has been anonymized by the same model that is being evaluated. The training data is LibriSpeech train-clean-360.\footnote{We experimented with finetuning this ASV model on the anonymized SEAME train data but did not observe an improvement compared to the LibriSpeech-only ASV model.} The privacy protection is then measured as the equal error rate (EER) of the ASV, in which a higher EER signifies better privacy. We perform ASV for female and male speakers separately and report their average. The performance is also assessed on the original, non-anonymized data, for which the ASV model is trained on the non-anonymized train data. The utility is typically measured as the word error rate (WER) of an English ASR system. We extend this to the mixed error rate (MER), which corresponds to WER for English and Spanish words, and to the character error rate (CER) for Mandarin ones. Following \cite{meyer24_multilingual}, we use Whisper-large-v3\footnote{\url{https://huggingface.co/openai/whisper-large-v3}} for this purpose, similar to its use in the \our{} model.  Before computing the MER, we convert Chinese characters to Pinyin. Additionally, we measure the phone error rate (PER) using the phonemizer of the IMS Toucan toolkit \cite{toucan}.

\subsection{Recognizing code-switching speech with Whisper}
Whisper is used for ASR in the anonymization of \our{} as well as the utility evaluation. It is currently one of the most powerful multilingual ASR systems but is not specifically trained for code-switching. In fact, the model expects a single language per audio. However, we observe in our experiments that Whisper is able to transcribe code-switched speech, and is thus suitable for preliminary investigations of this topic, given that other code-switching specific ASR models like \cite{yang2024adapting} are not publicly accessible. We experiment with different language prompts to find the optimal settings for the two datasets. For \seame{}, we achieve the best results by prompting Whisper with the language \textit{Mandarin}. For \miami{}, it is best to let Whisper recognize the language itself, but we rerun the recognition using the prompt \textit{Spanish} if the recognized language is different from Spanish and English. Moreover, we set the output probabilities of special characters (e.g., digits, currency signs) to a low value such that Whisper would transcribe these words as pronounced. We use the same settings for the Whisper version in \our{} and the utility evaluation.

\section{Experiments}
\begin{table}[t]
    \centering
    \caption{Anonymization on VPC datasets. For WER and original EER, lower is better. For anonymized EER, higher is better.}
    \begin{tabular}{c|rr|rr}
        \toprule
                           & \multicolumn{2}{c|}{EER (\%)}        & \multicolumn{2}{c}{WER (\%)} \\
         Anon Model        & Libri          & VCTK           & Libri          & VCTK \\
         \midrule
         \textit{Original} & \textit{4.59}  & \textit{2.37}  & \textit{2.77}  & \textit{1.57} \\
         \mcadams{}          &  7.81          &  4.69          &  4.13          &  9.83 \\
         \xiaoxiao{}         &  2.05          &  3.46          &  \textbf{3.39} &  3.11 \\
         \our{}              & \textbf{46.60} & \textbf{50.01} &  3.64          &  \textbf{2.24} \\
         \bottomrule
    \end{tabular}
    \label{tab:sanity_check}
\end{table}

\begin{table*}[t]
    \centering
    \caption{Anonymization on code-switching datasets. For reference, all scores are also computed for the original data (Orig).}
    \begin{tabular}{cc||rrrr||rrrr|rrrr}
        \toprule
                                 &  & \multicolumn{4}{c||}{Privacy}                        & \multicolumn{8}{c}{Utility} \\
                                 &  & \multicolumn{4}{c||}{EER (\%) $\uparrow$}                                        & \multicolumn{4}{c|}{MER (\%) $\downarrow$}       & \multicolumn{4}{c}{PER (\%) $\downarrow$} \\
                                 &  & \textit{Orig}$\downarrow$   & \mcadams   & \xiaoxiao{}  & \our{}            & \textit{Orig}  & \mcadams   & \xiaoxiao{}   & \our{}           & \textit{Orig}  & \mcadams   & \xiaoxiao{}   & \our{}\\
         \midrule
         \multirow{3}{*}{\seame{}} & EN & \textit{7.54} & 9.46  & 7.41 & \textbf{48.93} & \textit{27.39} & 57.05 & 41.41 & \textbf{28.26} & \textit{16.86} & 40.62 & 27.02 & \textbf{17.29}\\
                                 & ZH & \textit{6.23} & 9.45  & 7.59 & \textbf{48.40} & \textit{19.36} & 44.05 & 40.28 & \textbf{25.31} & \textit{14.55} & 33.75 & 29.57 & \textbf{17.61}\\
                                 & CS & \textit{3.18} & 4.49  & 3.72 & \textbf{50.61} & \textit{38.72} & 56.97 & 57.22 & \textbf{44.27} & \textit{30.64} & 44.58 & 43.53 & \textbf{35.05}\\
        \midrule
        \multirow{3}{*}{\miami{}}  & EN & \textit{8.48} & 21.77 & 6.29 & \textbf{49.22} & \textit{20.12} & 52.13 & 33.16 & \textbf{21.28} & \textit{15.03} & 41.08 & 25.31 & \textbf{15.26}\\
                                 & ES & \textit{8.78} & 19.81 & 7.08 & \textbf{48.61} & \textit{28.86} & 69.37 & 70.89 & \textbf{30.56} & \textit{20.71} & 52.37 & 51.16 & \textbf{21.62}\\
                                 & CS & \textit{6.75} & 20.81 & 4.19 & \textbf{46.94} & \textit{39.29} & 74.28 & 78.59 & \textbf{43.02} & \textit{30.39} & 55.59 & 58.41 & \textbf{32.39}\\
        \bottomrule
    \end{tabular}
    \label{tab:anonymization_results}
\end{table*}

\subsection{Anonymization of VPC datasets}

In order to understand the general anonymization abilities of each VA model, we first test them on the English datasets provided by the VPC \cite{vpc22_taslp}. The results are shown in Table \ref{tab:sanity_check}. 
All methods increase privacy protection except \xiaoxiao{} on LibriSpeech. 
Generally, the anonymization of \mcadams{} and \xiaoxiao{} result in only weak protection of privacy, with EER scores below 10\%. In contrast, the results of \our{} are close to 50\% EER, marking a strong anonymization. It also achieves the best utility preservation, having on average the lowest MER scores. 
We note that the performance of \xiaoxiao{} is lower than  reported in \cite{miao_2022_language, miao_2022_analyzing} because they used a less informed evaluation strategy of earlier VPC editions. On the other hand, our privacy results for \our{} are higher than in \cite{meyer24_multilingual}, which is mainly due to disabling the prosody cloning mechanism, leading to less leakage of speaker information during VA.

\subsection{Anonymization of code-switching data}

The results of the VA on \seame{} and \miami{} are shown in Table \ref{tab:anonymization_results}. In terms of privacy, we see the same trends across anonymization systems as for LibriSpeech and VCTK. The data of \mcadams{} and \xiaoxiao{} can be identified almost as well as the original data, with EER scores below 10\%. 
Interestingly, this trend is broken for \mcadams{} on the Miami dataset: There, the privacy protection is considerably higher, with EER between 20 and 26\%, which is due to increased noise levels after VA. For \our{}, the results are equally good across all datasets, with EER close to 50\% suggesting an overall successful anonymization. 

Similarly, the \our{} model is better at preserving the utility of the speech. There is an absolute degradation of 1-6\% MER across datasets, but less pronounced than the 20-41\% MER of \mcadams{} and 13-42\% MER of \xiaoxiao{}. Both of them distort the speech in such a way that utility is heavily affected. For \mcadams{}, the degradation remains relatively constant across datasets and language settings. For \xiaoxiao{}, though being considered language-independent, the anonymization has a smaller effect on English utterances than Spanish and code-switching speech. The PER scores are generally lower but show the same differences between the models, thus confirming these trends.

The differences between code-switching and monolingual utterances can be seen by comparing their results on original speech. Speaker verification seems to be facilitated by code-switching, leading to EER scores that are 2-4 points lower than on monolingual data. 
For utility, on the other hand, the ASR has more difficulties in recognizing the speech in CS than the monolingual subsets, with an absolute difference of up to 20\% MER. 
However, given that Whisper expects audios to be monolingual, the error rate is with less than 40\% MER still relatively small. We will analyze this finding in the following.

\section{Analysis}

Among the VA systems we tested, only \our{} could achieve a high level of privacy and keep the utility degradation comparably low. However, the utility scores on the original data are substantially higher than on the standard VPC data. Thus, we examine the outcome of the ASR before and after anonymization with \our{} more closely to understand to what extent the code-switching behavior is kept and reflected in the results.

\subsection{Preservation of code-switching during anonymization}
\begin{table}[h!]
    \centering
    \caption{Code-switching points before (CSP(O)) and after (CSP(A)) anonymization. * denotes a significant difference compared to the total MER.}
    \resizebox{\columnwidth}{!}{
    \begin{tabular}{l|ll|ll}
        \toprule
         & \multicolumn{2}{c|}{\# utterances} & \multicolumn{2}{c}{MER} \\
         & \seame{} & \miami{} & \seame{} & \miami{} \\
        \midrule
        Total & 7,533 & 536 & 44.24 & 43.90 \\
        $\text{CSP}(\text{A}) <  \text{CSP}(\text{O})$ & 4,926 (65\%) & 382 (71\%) & 50.90* & 50.58* \\
        $\text{CSP}(\text{A}) = 0 $ & 2,281 (30\%) & 342 (64\%) & 59.36* & 51.95*\\
        $\text{CSP}(\text{A}) = \text{CSP}(\text{O})$ & 2,232 (30\%) & 137 (26\%) & 27.81* & 24.38*\\
        \bottomrule
    \end{tabular}}
    \label{tab:csp}
\end{table}

We compare the number of code-switching points of an utterance in CS before (CSP(O)) and after (CSP(A)) anonymization with \our{} in order to estimate how much the model changes the code-switching behavior.
Although this does not show if the code-switching is kept for the same words but only the frequency, it gives some indication of how the model deals with code-switching. 
The transcription-based language identification tools of \our{} are reused to map each recognized token to a language and count the number of language changes in an utterance. We notice $\text{CSP(O)} = 0$ for 0.3\% of utterances in \seame{} and 14\% in \miami{}, even though they had been annotated as code-switching. 
Thus, we only use utterances with $\text{CSP(O)} > 0$ in our analysis,  shown in Table \ref{tab:csp}. On average, code-switching in \seame{} is reduced from $\text{CSP(O)}=3.47$ to $\text{CSP(A)}=1.75$, with less code-switching points for 65\% and no code-switching for 30\% of utterances after VA. In a similar 30\% of cases, the code-switching remains stable ($\text{CSP}(\text{A}) = \text{CSP}(\text{O})$). For \miami{}, the reduction is more severe: 71\% of utterances have less and 64\% no code-switching after anonymization, it stays the same in 26\% of cases. On average, code-switching is reduced from $\text{CSP(O)}=1.47$ to $\text{CSP(A)}=0.52$. This affects the MER: For both \seame{} and \miami{}, the MER is significantly\footnote{All statistical significance tests were performed with the one-sided Mann-Whitney U test and $\alpha=0.025$.} higher if $\text{CSP(O)} = 0$ and significantly lower if $\text{CSP}(\text{A}) = \text{CSP}(\text{O})$, as compared to the total MER. 
The fact that the anonymized utterances have in most cases less code-switching than before might explain the higher total MER for CS compared to the monolingual subsets.

\subsection{Subjective analysis of anonymized speech content}
To understand how the speech content is changed during VA, we perform a small user study with 6 subjects each for \seame{} and \miami{}, selected based on their knowledge of the respective languages. For each dataset, we present ten CS audios as anonymized with \our{} and their gold transcription to the subjects and ask them to compare the content of the audio to the text. The answers reveal how the errors made by the ASR in \our{} are reflected in the anonymized speech. In both datasets, parts of one language were regularly translated into the more dominant language or recognized as similar-sounding words (near-homophones) of the other language. For \miami{}, the recognition would ignore one language altogether in some cases. Several words would be recognized as near-homophones in the same language. This same-language change was not always perceived as such by the listeners, e.g., 4 out of 6 did not hear the difference between ``jobs" and ``jumps" (\miami{}). Interestingly, for two utterances in \seame{}, the content of the anonymized audio almost completely matched with the gold transcript but was mistranscribed during MER computation, resulting in an unfairly high error rate. The outcome of the user study raises the question of which kind of utterances might be more prone to transcription errors than others.

\subsection{ASR performance based on data characteristics}
Since \seame{} and \miami{} come with rich annotations, we can divide the dataset into subsets depending on the presence of certain phenomena in an utterance. By comparing the MER achieved on each dataset after removing this subset to the scores in Table \ref{tab:anonymization_results}, we can estimate the effect of this characteristic on the utility computation. We report the results on the original data but could observe the same trends also after VA. In this analysis, we find that utterances containing abbreviations and acronyms (e.g., USA), foreign words in languages other than the two dataset languages, and filled pauses and discourse particles (e.g., ``oh"), have a significant impact on the MER in \seame{}. Without the respective samples, the dataset has only 29\% (CS) to 38\% (EN) of its previous size left but the MER is also reduced by 24\% (EN, 21\% MER), 14\% (ZH, 17\% MER) and 11\% (CS, 35\% MER). This shows a higher influence of these characteristics for monolingual than code-switching speech. For \miami{}, the most influential phenomena are the presence of words tagged as \textit{unknown} by the annotator, repeated words or phrases, trailing off and incomplete utterances, and audios with a sub-average loudness. Their removal leads to a dataset of 37\% (CS) to 47\% (ES) of its original size and reduces the MER by 17\% for EN (17\% MER) and 15\% for ES (24\% MER). In contrast, the MER increases by 8\% for CS (42\% MER). 
We conclude that the characteristics that lead to errors in speech recognition are different for code-switching than for monolingual utterances. To improve the performance for CS, code-switching-specific processing is necessary.

\section{Conclusion}
In this paper, we present the first work of applying voice anonymization to code-switching speech. We prepare two  Mandarin-English and Spanish-English datasets for the evaluation and experiment with three anonymization methods. Two of them are designed as language-independent but fail to achieve sufficient privacy preservation even on monolingual data. The third method is a multilingual model which we adapt for code-switching and which performs well in terms of privacy and utility preservation. We find, however, that Whisper, used for utility evaluation and in the multilingual model, has several difficulties in recognizing the speech in these datasets.
In future work, further investigations should be made with other evaluation models and aspects such as speech emotion recognition, as well as more code-switching pairs and datasets.

\section{Acknowledgements}

This work is funded by the Deutsche Forschungsgemeinschaft (DFG, German Research  Foundation) – Project: Multilingual Controllable Voice Privacy (VoiPy) - Project number 533241795.

\bibliographystyle{IEEEtran}
\bibliography{literature}

\end{document}